\documentclass[universe,article,accept,moreauthors,pdftex,10pt,a4paper]{Definitions/mdpi} 
\usepackage[utf8]{inputenc}
\usepackage{booktabs} 
\usepackage{multirow}
\usepackage{soul} 
\usepackage{microtype}

\firstpage{1} 
\pubvolume{5}
\issuenum{2}
\articlenumber{42}
\pubyear{2019}
\copyrightyear{2019}
\history{Received: 29 November 2018; Accepted: 16 January 2019; Published: 23 January 2019}
\issuenum{2}
 \updates{yes}

\Title{QUBIC: Exploring the Primordial Universe with the Q\&U Bolometric Interferometer}

\Author{
Aniello	Mennella	$^{1,2,}$*, Peter	Ade	$^{3}$, Giorgio	Amico	$^{4}$, Didier	Auguste	$^{5}$, Jonathan	Aumont	$^{6}$, Stefano	Banfi	$^{7}$, Gustavo	Barbaràn	$^{8}$, Paola	Battaglia	$^{9}$, Elia	Battistelli	$^{4,10}$, Alessandro	Baù	$^{7,11}$
% ...
}

\AuthorNames{
Aniello	Mennella,	
Peter	Ade	,
Giorgio	Amico	,
Didier	Auguste	,
Jonathan	Aumont	,
Stefano	Banfi	,
Gustavo	Barbaràn	,
Paola	Battaglia	,
Elia	Battistelli	,
Alessandro	Baù	,
Benoit	Bélier	,
David G.	Bennett	,
Laurent	Bergé	,
Jean Philippe	Bernard	,
Marco	Bersanelli	,
Marie Anne	Bigot Sazy	,
Nathat	Bleurvacq	,
Juan	Bonaparte	,
Julien	Bonis	,
Emory F.	Bunn	,
David Burke, 
Daniele	Buzi	,
Alessandro	Buzzelli	,
Francesco	Cavaliere	,
Pierre	Chanial	,
Claude	Chapron	,
Romain	Charlassier	,
Fabio	Columbro	,
Gabriele	Coppi	,
Alessandro	Coppolecchia	,
Rocco	D'Agostino	,
Giuseppe	D'Alessandro	,
Paolo	De Bernardis	,
Giancarlo	De Gasperis	,
Michele	De Leo	,
Marco	De Petris	,
Andres	Di Donato	,
Louis	Dumoulin	,
Alberto	Etchegoyen	,
Adrián	Fasciszewski	,
Cristian	Franceschet	,
Martin Miguel	Gamboa Lerena	,
Beatriz	Garcia	,
Xavier	Garrido	,
Michel	Gaspard	,
Amanda	Gault	,
Donnacha	Gayer	,
Massimo	Gervasi	,
Martin	Giard	,
Yannick Giraud	Héraud	,
Mariano	Gómez Berisso	,
Manuel	González	,
Marcin	Gradziel	,
Laurent	Grandsire	,
Eric	Guerard	,
Jean Christophe	Hamilton	,
Diego	Harari	,
Vic	Haynes	,
Sophie	Henrot Versillé	,
Duc Thuong	Hoang	,
Nicolas	Holtzer	,
Federico	Incardona	,
Eric	Jules	,
Jean	Kaplan	,
Andrei	Korotkov	,
Christian	Kristukat	,
Luca	Lamagna	,
Sotiris	Loucatos	,
Amy Lowitz,
Vladimir	Lukovic	,
Louis	Thibault	,
Ra\`ul Horacio	Luterstein	,
Bruno	Maffei	,
Stefanos	Marnieros	,
Silvia	Masi	,
Angelo	Mattei	,
Andrew	May	,
Mark	McCulloch	,
Maria C.	Medina	,
Lorenzo	Mele	,
Simon J.	Melhuish	,
Ludovic	Montier	,
Louise Mousset,
Luis Mariano	Mundo	,
John Anthony	Murphy	,
James	Murphy	,
Creidhe	O'Sullivan	,
Emiliano	Olivieri	,
Alessandro	Paiella	,
Francois	Pajot	,
Andrea	Passerini	,
Hernan	Pastoriza	,
Alessandro	Pelosi	,
Camille	Perbost	,
Maurizio Perciballi, 
Federico	Pezzotta	,
Francesco	Piacentini	,
Michel	Piat	,
Lucio	Piccirillo	,
Giampaolo	Pisano	,
Gianluca	Polenta	,
Damien	Prêle	,
Roberto	Puddu	,
Damien	Rambaud	,
Pablo	Ringegni	,
Gustavo E.	Romero	,
Maria	Salatino	,
Alessandro	Schillaci	,
Claudia G.	Scóccola	,
Stephen P.	Scully	,
Sebastiano	Spinelli	,
Michail	Stolpovskiy	,
Federico	Suarez	,
Guillaume Stankowiak,
Andrea	Tartari	,
Jean Pierre	Thermeau	,
Peter	Timbie	,
Maurizio Tomasi,
Steve A.	Torchinsky	,
Mathieu	Tristram	,
Gregory S.	Tucker	,
Carole E.	Tucker	,
Sylvain	Vanneste	,
Daniele	Viganò	,
Nicola	Vittorio	,
Fabrice	Voisin	,
Robert	Watson	,
Francois	Wicek	,
Mario	Zannoni	and
Antonio	Zullo	
}

\address{%
$^{1}$\quad	\hspace{.8mm}Department of Physics, University of Milan,  20133~Milano, Italy; marco.bersanelli@fisica.unimi.it (M.B.); francesco.cavaliere@unimi.it (F.C.); cristian.franceschet@fisica.unimi.it (C.F.); federico.incardona@unimi.it~(F.I.); federico.pezzotta@unimi.it (F.P.); maurizio.tomasi@unimi.it (M.T.); daniele.vigano@unimi.it (D.V.)\\
$^{2}$\quad	\hspace{.8mm}Istituto Nazionale di Fisica Nucleare Milano 1 Section, 20133~Milano, Italy\\
$^{3}$\quad	\hspace{.8mm}School of Physics and Astronomu, Cardiff University, Cardiff CF10 3AT, UK; adepa@cardiff.ac.uk (P.A.); giampaolo.pisano@astro.cf.ac.uk (G.P.); carole.tucker@astro.cf.ac.uk (C.E.T.)\\
$^{4}$\quad	\hspace{.8mm}Department of Physics, Universit\`a di Roma La Sapienza, 00185~Roma, Italy; giorgio.amico@uniroma1.it~(G.A.); Elia.Battistelli@roma1.infn.it (E.B.); daniele.buzi.db@gmail.com (D.B.); \newline fabio.columbro@roma1.infn.it (F.C.); alessandro.coppolecchia@roma1.infn.it (A.C.); giuseppe.dalessandro@roma1.infn.it (G.D.); Paolo.DeBernardis@roma1.infn.it (P.D.B.); \newline k-9@libero.it (M.D.L.); Marco.DePetris@roma1.infn.it (M.D.P.); luca.lamagna@roma1.infn.it (L.L.); Silvia.masi@roma1.infn.it (S.M.); lmele64@gmail.com (L.M.); Alessandro.Paiella@roma1.infn.it (A.P.); Francesco.Piacentini@roma1.infn.it (F.P.); roberto.puddu@roma1.infn.it (R.P.); alex78@caltech.edu (A.S.); antonio.zullo@roma1.infn.it (A.Z.)\\
$^{5}$\quad	\hspace{.8mm}Laboratoire de l'Acc\'{e}l\'{e}rateur Lin\'{e}aire (CNRS-IN2P3), 91898~Orsay, France; auguste@lal.in2p3.fr (D.A.); bonis@lal.in2p3.fr (J.B.); garrido@lal.in2p3.fr (X.G.); gaspard@lal.in2p3.fr (M.G.); guerard@lal.in2p3.fr (E.G.); versille@lal.in2p3.fr (S.H.V.); jules@lal.in2p3.fr (E.J.); \newline louis@lal.in2p3.fr (T.L.); tristram@lal.in2p3.fr (M.T.); vanneste@lal.in2p3.fr (S.V.); \newline wicek@lal.in2p3.fr (F.W.)\\
$^{6}$\quad	\hspace{.8mm}Institut d'Astrophysique Spatiale (CNRS-INSU), 91405~Orsay, France; jonathan.aumont@ias.u-psud.fr (J.A.); Bruno.Maffei@ias.u-psud.fr (B.M.)\\
$^{7}$\quad	\hspace{.8mm}Department of Physics, Universit\`a di Milano Bicocca, 20126~Milano, Italy; stefano.banfi@mib.infn.it (S.B.); Alessandro.Bau@mib.infn.it (A.B.); Massimo.Gervasi@mib.infn.it (M.G.); Andrea.Passerini@mib.infn.it~(A.P.); sebspin@tiscali.it (S.S.); Mario.Zannoni@unimib.it (M.Z.)\\
$^{8}$\quad	\hspace{.8mm}Comisión Nacional De Energia Atómica, Salta A4400, Argentina; barbaran@cnea.gov.ar (G.B.); \newline bonaparte@cnea.gov.ar (J.B.); andresdidonato@cnea.gov.ar (A.D.D.); afascisz@cnea.gov.ar (A.F.); luter@cnea.gov.ar (R.H.L.)\\
$^{9}$\quad	\hspace{.8mm}Istituto Nazionale di Astrofisica/OAS Bologna, 40129~Bologna, Italy; paola.battaglia@inaf.it\\
$^{10}$\quad \hspace{-.2mm}Istituto Nazionale di Fisica Nucleare Roma 1 Section, 00185~Roma, Italy; angelo.mattei@roma1.infn.it (A.M.); alessandro.pelosi@roma1.infn.it (A.P.); maurizio.perciballi@roma1.infn.it (M.P.)\\
$^{11}$\quad	 \hspace{-.2mm}Istituto Nazionale di Fisica Nucleare Milano Bicocca Section, 20126~Milano, Italy\\
$^{	12	}$\quad	\hspace{-.2mm}Centre de Nanosciences et de Nanotechnologies, 91120~Palaiseau, France; benoit.belier@u-psud.fr\\
$^{	13	}$\quad	\hspace{-.2mm}Department of Experimental Physics, National University of Ireland, Mariavilla, Maynooth 99MX+QH, Ireland; david.g.bennett10@gmail.com (D.G.B.); david.burke.2012@mumail.ie (D.B.); donnacha.gayer.2009@mumail.ie (D.G.); marcin.gradziel@nuim.ie (M.G.); anthony.murphy@mu.ie (J.A.M.); james.murphy.2018@mumail.ie (J.D.M.); creidhe.osullivan@mu.ie (C.O.); stephen.scully@itcarlow.ie (S.P.S.)\\
$^{	14	}$\quad	\hspace{-.3mm}Centre de Spectrom\'{e}trie Nucl\'eaire et de Spectrom\'{e}trie de Masse (CNRS-IN2P3), 91405~Orsay, France; Laurent.Berge@csnsm.in2p3.fr (L.B.); Dumoulin@csnsm.in2p3.fr (L.D.);\newline nicholas.holtzer@csnsm.in2p3.fr (N.H.); Stefanos.Marnieros@csnsm.in2p3.fr (S.M.); Emiliano.Olivieri@csnsm.in2p3.fr (E.O.)\\
$^{	15	}$\quad	\hspace{-.3mm}Institut de Recherche en Astrophysique et Plan\'{e}tologie (CNRS-INSU), 31028~Toulouse, France; Jean-Philippe.Bernard@cesr.fr (J.P.B.); Giard@cesr.fr (M.G.); Ludovic.Montier@irap.omp.eu (L.M.); francois.pajot@irap.omp.eu (F.P.); Damien.Rambaud@irap.omp.eu (D.R.)\\
$^{	16	}$\quad	\hspace{-.3mm}AstroParticule et Cosmologie (CNRS-IN2P3), 75013~Paris, France; marieanne.bigotsazy@gmail.com~(M.A.B.S.); bleurvac@apc.univ-paris7.fr (N.B.); chanial@apc.univ-paris7.fr~(P.C.); chapron@apc.univ-paris7.fr (C.C.); romain.charlassier@turing-capital.com~(R.C.); Yannick.Giraud-Heraud@apc.univ-paris-diderot.fr (Y.G.H.); lgrandsire@apc.in2p3.fr (L.G.); Hamilton@apc.in2p3.fr (J.C.H.); hoang@apc.in2p3.fr (D.T.H.); kaplan@apc.univ-paris7.fr (J.K.); loucatos@apc.univ-paris7.fr (S.L.); mousset@apc.in2p3.fr (L.M.); camille.perbost@apc.univ-paris7.fr (C.P.); piat@apc.univ-paris7.fr (M.P.); prele@apc.in2p3.fr (D.P.); salatino@apc.in2p3.fr (M.S.); guillaume.stankowiak@apc.univ-paris7.fr (G.S.); mikhail.stolpovskiy@apc.univ-paris7.fr (M.S.); jean-pierre.thermeau@univ-paris-diderot.fr (J.P.T.); satorchi@apc.in2p3.fr (S.A.T.); voisin@apc.in2p3.fr (F.V.)\\
$^{	17	}$\quad	\hspace{-.2mm}Department of Physics, Richmond University, Richmond, VA 23173, USA; ebunn@richmond.edu\\
$^{	18	}$\quad	\hspace{-.2mm}Dipartimento di Fisica, Universit\`a di Roma Tor Vergata, 00133~Roma, Italy; alessandro.buzzelli@roma2.infn.it (A.B.); rocco.dagostino@roma2.infn.it (R.D.); giancarlo.degasperis@roma2.infn.it (G.D.G.); vladimir.lukovic@roma2.infn.it (V.L.); nicola.vittorio@uniroma2.it (N.V.)\\
$^{	19	}$\quad	\hspace{-.2mm}Istituto Nazionale di Fisica Nucleare Roma Tor Vergata section, 00133~Roma, Italy\\
$^{	20	}$\quad	\hspace{-.2mm}School of Physics \& Astronomy, University of Manchester,  Manchester M13 9PL, UK; gabriele.coppi@postgrad.manchester.ac.uk (G.C.); Vhaynes@jb.man.ac.uk (V.H.); andrew.may-3@postgrad.manchester.ac.uk (A.M.); mark.mcculloch@manchester.ac.uk (M.M.); Simon.Melhuish@manchester.ac.uk  (S.J.M.); lucio@jb.man.ac.uk (L.P.); Bob.Watson@manchester.ac.uk (B.W.)\\
$^{	21	}$\quad	\hspace{-.2mm}Department of Physics, University of Surrey, Guildford GU2 7XH, UK; m.deleo@surrey.ac.uk\\
$^{	22	}$\quad	\hspace{-.2mm}Instituto de Tecnologías en Detección y Astropartículas, Buenos Aires B1650, Argentina; alberto.etchegoyen@iteda.cnea.gov.ar (A.E.); beatriz.garcia@iteda.cnea.gov.ar (B.G.); federico.suarez@iteda.cnea.gov.ar (F.S.)\\
$^{	23	}$\quad	\hspace{-.2mm}Facultad de Ciencias Astronómicas y Geofísicas, Univ. Nacional de la Plata, La Plata B1900FWA, Argentina; mgamboa@fcaglp.unlp.edu.ar (M.M.G.L.); luis.mundo@ing.unlp.edu.ar (L.M.M.); ringegni@ing.unlp.edu.ar~(P.R.); cscoccola@fcaglp.unlp.edu.ar (C.G.S.)\\
$^{	24	}$\quad	\hspace{-.2mm}Department of Physics, University of Wisconsin, Madison, WI 53706, USA; amanda@physics.wisc.edu~(A.G.);  lowitz@wisc.edu (A.L.); pttimbie@wisc.edu (P.T.)\\
$^{	25	}$\quad	\hspace{-.2mm}Ctr. Atómico Bariloche y Instituto Balseiro, CNEA, San Carlos de Bariloche R8402AGP, Argentina; berisso@cab.cnea.gov.ar (M.G.B.); manuel.gonzalez@ib.edu.ar (M.G.); diego.harari@gmail.com (D.H.); hernan@cab.cnea.gov.ar (H.P.)\\
$^{	26	}$\quad	\hspace{-.2mm}Faculty of Physics, University of Science and Technology of Hanoi (USTH), Vietnam Academy of Science and Technology (VAST), Ha Noi 10000, Vietnam\\
$^{	27	}$\quad	\hspace{-.2mm}Department of Physics, Brown University, Providence, RI~02912, USA; andrei\_korotkov@brown.edu (A.K.); Gregory\_Tucker@brown.edu (G.S.T.)\\
$^{	28	}$\quad	\hspace{-.2mm}Escuela de Ciencia y Tecnología, Universidad Nacional de San Martín, San Martin 1650, Argentina; kristukat@cnea.gov.ar\\
$^{	29	}$\quad	\hspace{-.2mm}Instituto Argentino de Radioastronomía, Berazategui 1880, Argentina; clementina.medina@gmail.com~(M.C.M.); gustavo.esteban.romero@gmail.com (G.E.R.)\\
$^{	30	}$\quad	\hspace{-.2mm}Agenzia Spaziale Iitaliana, 00133~Rome, Italy; Gianluca.Polenta@asdc.asi.it\\
$^{	31	}$\quad	\hspace{-.2mm}Institute of Technology, Carlow R93 A003, Ireland\\
$^{	32	}$\quad	\hspace{-.2mm}Istituto Nazionale di Fisica Nucleare Pisa Section, 56127~Pisa, Italy; andrea.tartari@pi.infn.it
}

\corres{Correspondence: aniello.mennella@fisica.unimi.it}

\abstract{In this paper, we describe QUBIC
, an experiment that will observe the polarized microwave sky with a novel approach, which combines the sensitivity of state-of-the-art bolometric detectors with the systematic effects control typical of interferometers. QUBIC's unique features are the so-called ``self-calibration'', a technique that allows us to clean the measured data from instrumental effects, and its spectral imaging power, i.e., the ability to separate the signal into various sub-bands within each frequency band. QUBIC will observe the sky in two main frequency bands: 150\,GHz and 220\,GHz. A~technological demonstrator is currently under testing and will be deployed in Argentina during 2019, while the final instrument is expected to be installed during 2020.}


\begin{document}


\section{Introduction}
\label{sec_introduction}

The Q\&U Bolometric Interferometer for Cosmology (QUBIC)
is an experiment based on the concept of bolometric interferometry \cite{Battistelli2011} and designed to measure the $B$-mode polarization anisotropies of the Cosmic Microwave Background (CMB). The~QUBIC design combines the sensitivity of Transition Edge Sensors (TES) bolometric detectors with the systematic effects and foreground control provided by its interferometric design.

The control of astrophysical foregrounds, in particular, is a factor of increasing importance in CMB polarization experiments, and QUBIC allows us to disentangle sub-bands in each main frequency band thanks to its spectral imaging capability, which is deeply rooted in the interferometric nature of the~instrument.

QUBIC will operate from the ground observing the sky in two main spectral bands centered at 150~and 220\,GHz \cite{qubic15} and will be deployed in Argentina, at the Alto Chorrillos site. The team is currently finalizing the laboratory tests of the technical demonstrator, a simplified version of the instrument that will be installed at the site during 2019 and will demonstrate the technical and scientific potential of our approach. The final instrument will be deployed during 2020.

\section{The Instrument}

\begin{figure}[H]
\centering
  \includegraphics[width=15cm]{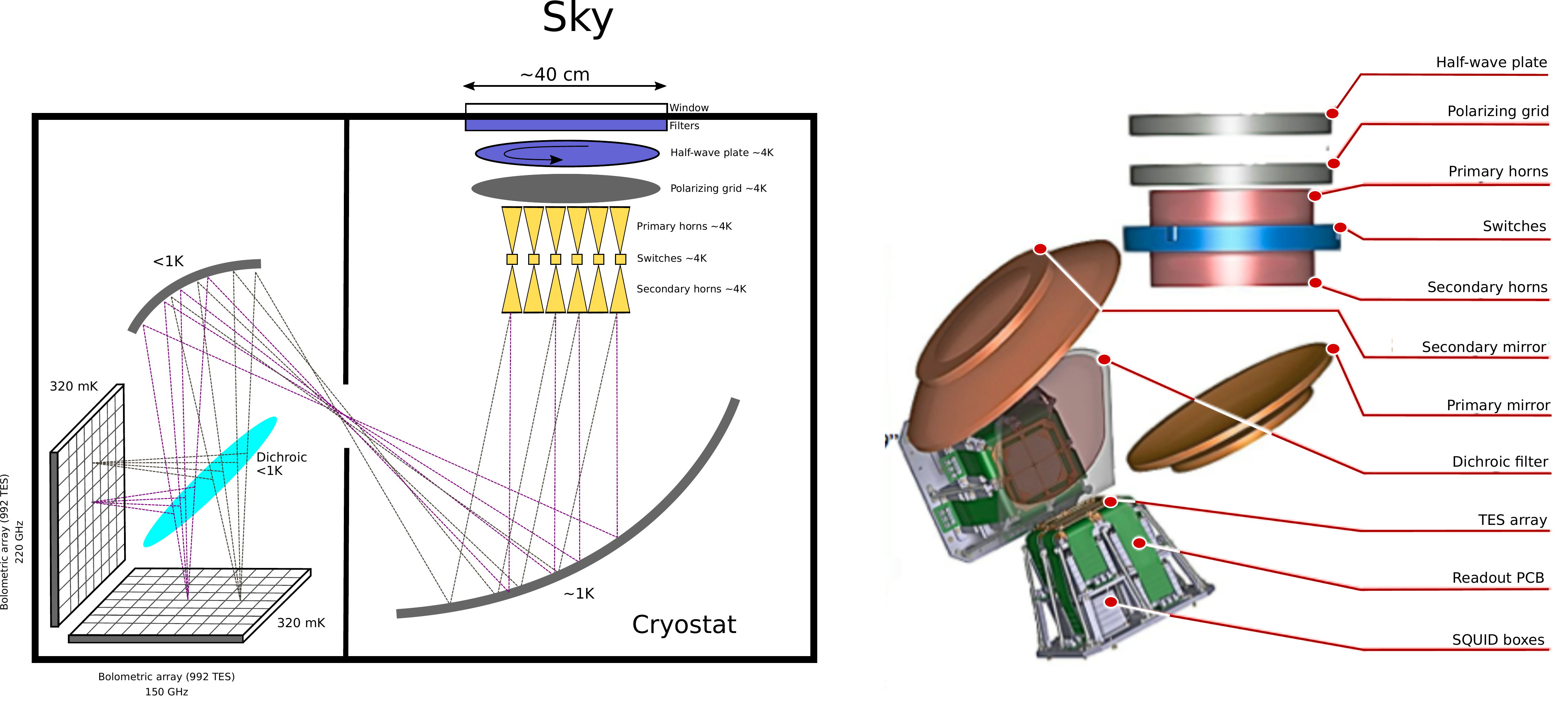}
  \caption{\label{fig_qubic_schematic}\textls[-15]{({\textbf{Left}}) Schematic of the QUBIC
	instrument. The window aperture is about 40\,cm; the cryostat is about 1.41\,m in diameter and 1.51\,m in height; ({\textbf{right}}) 3D rendering of the inner part of the cryostat. TES, Transition Edge Sensor.}}
\end{figure}

Figure~\ref{fig_qubic_schematic} shows a schematics of QUBIC. The signal from the sky enters the cryostat through a High-Density Polyethylene (HDPE) window.~Then, a rotating half-wave plate modulates the polarization, and a polarizing grid selects one of the two linear polarization components. An array of 400 back-to-back corrugated horns collects the radiation and re-images it onto a dual-mirror optical combiner that focuses the signal onto two orthogonal TES detectors focal planes. A dichroic filter placed between the optical combiner and the focal planes selects the two frequency bands, centered at 150 GHz and 220 GHz. The right panel of Figure~\ref{fig_qubic_schematic} shows a 3D rendering of the inner part of the~cryostat.

A key part of the instrument is an array of movable shutters placed between the primary and secondary feed-horn arrays. Each shutter acts as an RF switch (a blade that can slide into a smooth circular waveguide), which is used to exclude particular baselines when the instrument operates in calibrating mode. We call this particular calibration strategy ``self-calibration'', which is a key feature of the QUBIC systematic effects control. The interested reader can find the detailed description of all instrument parts in \cite{Aumont2016} and the theory of self-calibration in \cite{liu2010,bs13}.


\section{Measurement, Self-Calibration, and Spectral Imaging}
\label{sec_measurement}
\vspace{-6pt}

\subsection{Signal Model and Synthetic Beam}
\label{sec_signal_model}
 In QUBIC, the optical combiner focuses the radiation emitted by the secondary horns onto the two focal planes so that the image that forms on the detector arrays is the result of the interference arising from the sum of the fields radiated from each of the 400 apertures.

 Therefore, the signal measured at time $t$ by a detector $p$ on the focal plane is:
 \begin{equation}
	 R(p,\nu,t)=K\left[S_I(p,\nu)+\cos(4\,\phi_\mathrm{HWP}(t))\,S_Q(p,\nu)+\sin(4\,\phi_\mathrm{HWP}(t))\,S_U(p,\nu)\right],
   \label{eq_signal_model}
 \end{equation}
 where $\nu$ is the frequency, $\phi_\mathrm{HWP}$ is the angle of the half-wave plate at time $t$, and $K$ is an overall calibration constant that takes into account the efficiency of the optical chain. The three terms $S_{I,Q,U}$ in Equation~(\ref{eq_signal_model}) represent the sky signal in intensity and polarization convolved with the so-called {synthetic beam}.

\begin{figure}[H]
\centering
   \includegraphics[width=14.5cm]{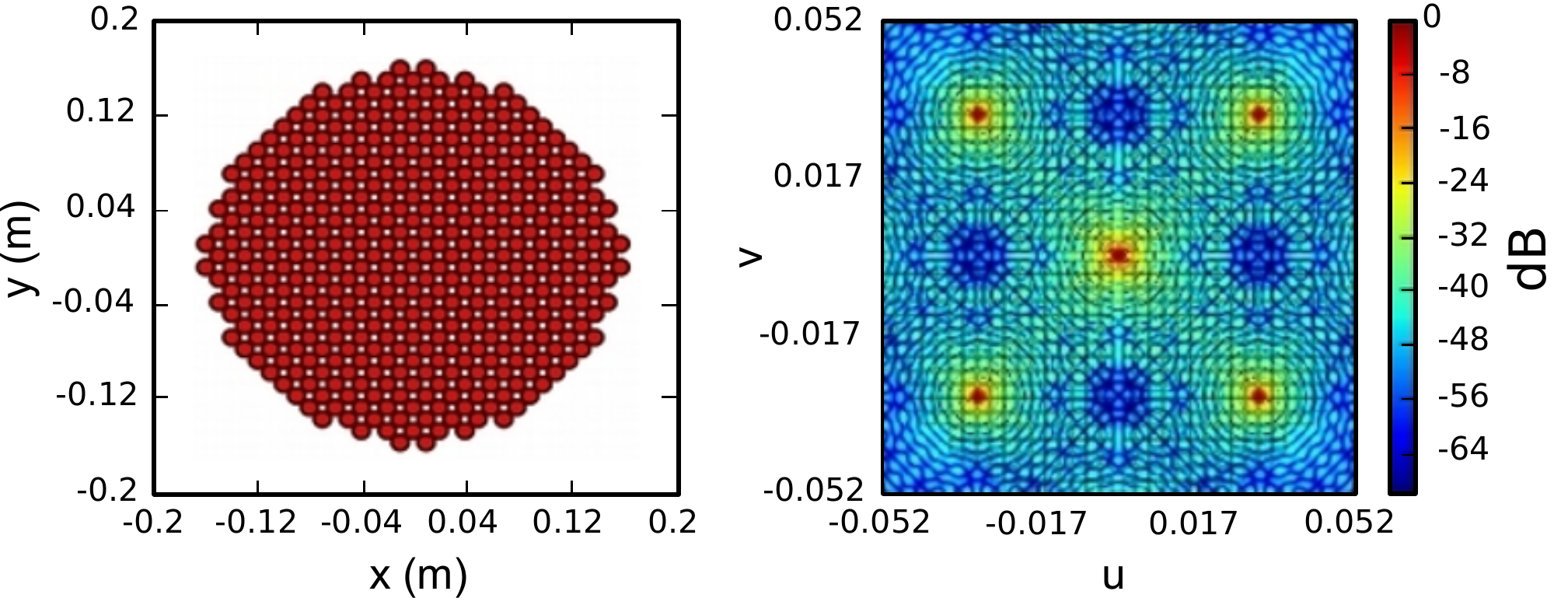}
   \caption{\label{fig_synthetic_beam}\textls[-15]{({\textbf{Left}}) QUBIC aperture plane showing all 400 antennas open to the sky; ({\textbf{right}}) the interference} pattern formed on each of the focal planes when the instrument is observing a point source located in the far field vertically along the instrument line-of-sight. The $u$ and $v$ coordinates are defined as: $u=\sin\theta\cos\phi$ and $v=\sin\theta\sin\phi$, where $\theta$ and $\phi$ are the angles on the celestial sphere defining the synthetic beam.}
\end{figure}

 The images in Figure~\ref{fig_synthetic_beam} help the reader to understand this concept qualitatively. Imagine that QUBIC observes a point source in the far field located directly along the line-of-sight with all 400 antennas open to the sky. The image formed on each of the focal planes (see the right panel of Figure~\ref{fig_synthetic_beam}) is an interference pattern formed by peaks and lobes. This pattern works like a {beam pattern} that convolves the sky signal.

 Therefore, if $X$ is the signal from the sky (either in intensity, $I$, or polarization, $Q,U$), then the measured signal on the pixel $p$ is $S_X(p) = \int X(\mathbf{n})B^p_\mathrm{synth}(\mathbf{n})d\mathbf{n}$. This means that QUBIC data can be analyzed similarly to the data obtained from a normal imager, provided that we build a window function of the synthetic pattern for each pixel. 

\subsection{Self-Calibration}
\label{sec_self_calibration}
  QUBIC {self-calibration} is a technique derived from radio-interferometry self-calibration \cite{liu2010}. In~QUBIC, self-calibration exploits the redundant interferometric patterns obtained when we selectively close various combinations of the 400 instrument apertures. 
  
\begin{figure}[H]
\centering
     \includegraphics[width=14cm]{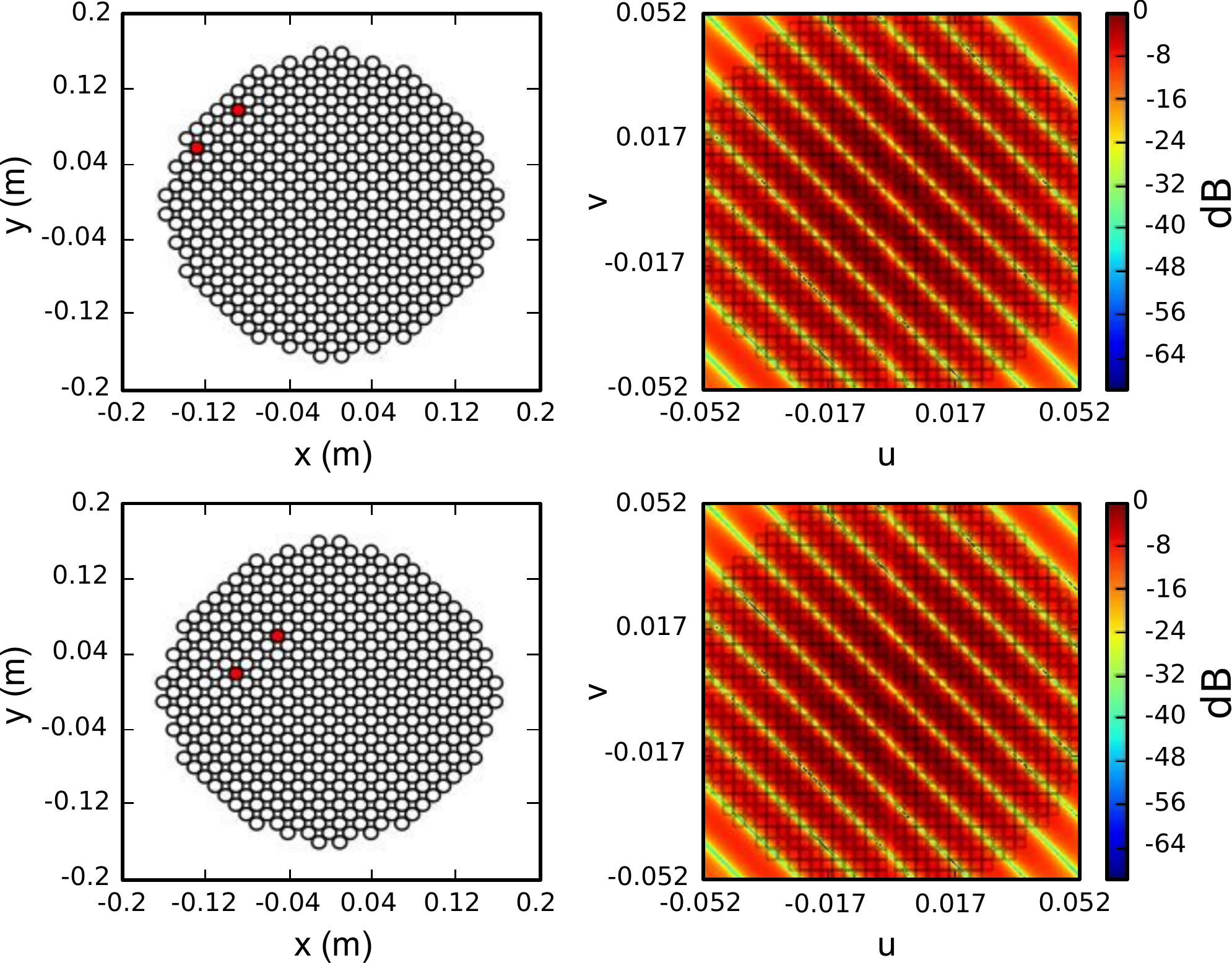}
   \caption{\label{fig_self_calibration_concept}Schematics of QUBIC self-calibration. The pictures in the two rows show that if we open any pair of horns with a given baseline, then we should ideally measure exactly the same interference~pattern.}
  \end{figure}

  To understand the basic concept, see the four panels of Figure~\ref{fig_self_calibration_concept}. The top-right plot shows the interference pattern arising from a horn configuration in which only two horns are open and all the others are closed (top-left). The panels in the bottom row show that if we open any other horn pair with the same baseline, we should ideally obtain exactly the same interference pattern. Furthermore, in~\cite{bs13}, Bigot-Sazy et al. showed that the configuration in which only two horns are open is equivalent to the complementary arrangement in which only two horns are closed.

  We can now use the fact that, for an ideal instrument, the interferometric pattern depends only on the baseline. This allows us to characterize the instrumental parameters using an~observation mode called self-calibration. During self-calibration, pairs of horns are successively shut while QUBIC observes an artificial partially-polarized source (a microwave synthesizer or a Gunn oscillator) in the far field. Then, we reconstruct the signal measured by each individual pair of horns in the array and compare them. 
  
  The point now is that if the source is stable and carefully monitored, then redundant baselines correspond to the same mode of the observed field, so that a different signal between them can only be due to photon noise or instrumental systematic effects. Using a detailed parametric model of the instrument, we can fully recover the instrument parameters through a non-linear inversion process. The updated model of the instrument can then be used to reconstruct the synthetic beam and improve the map-making, thus reducing the leakage from $E$- to $B$-modes.

\begin{figure}[H]
\centering
     \includegraphics[width=9cm]{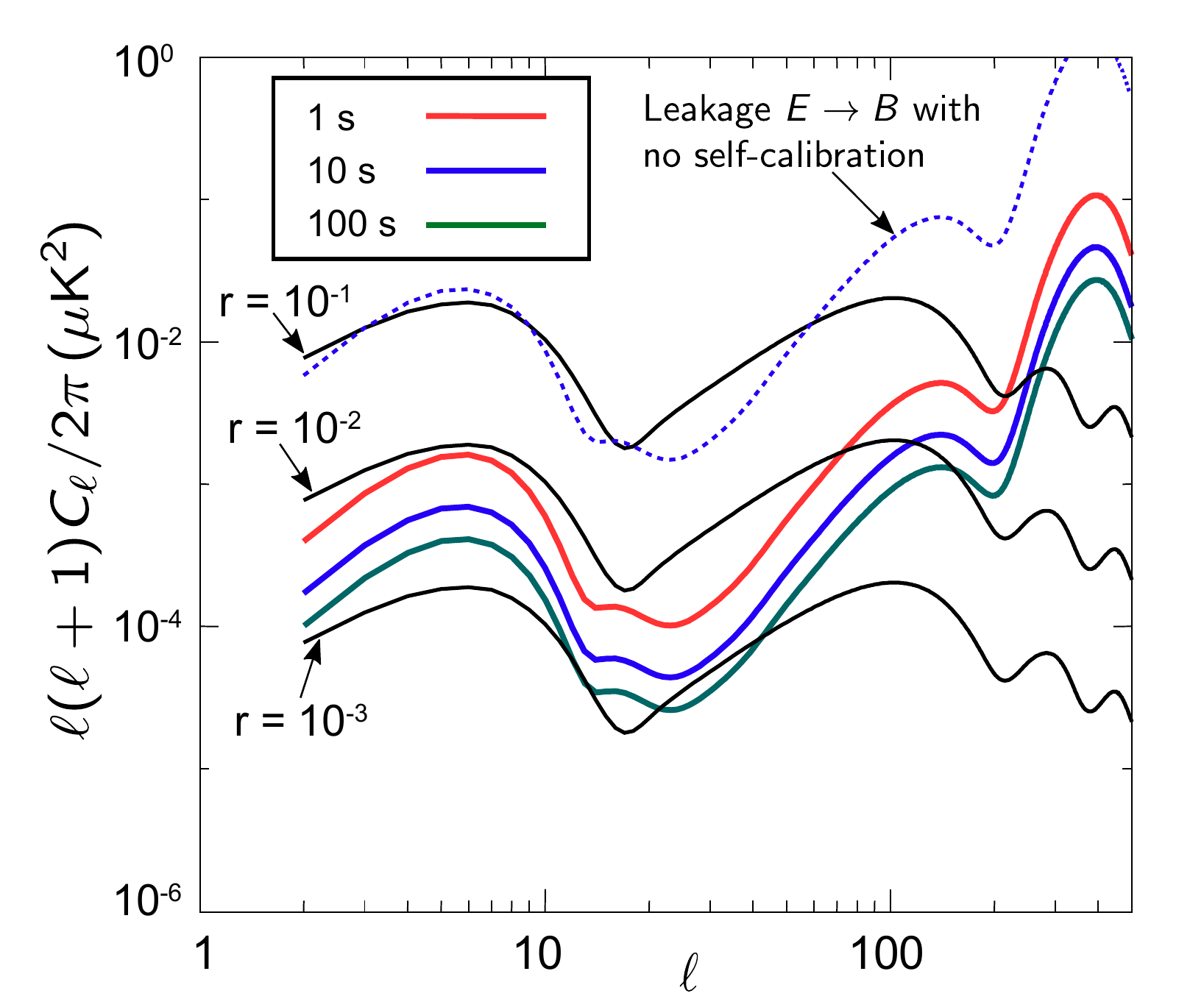}
     \caption{\label{fig_self_calibration_performance}Improvement in the recovery of the $B$-mode power spectrum as a function of the time spent in the self-calibration mode. The three curves drawn with black solid lines represent theoretical $B$-mode power spectra calculated for three different values of the tensor-to-scalar ratio, $r$.}
  \end{figure}

  In Figure~\ref{fig_self_calibration_performance} (adapted from \cite{bs13}), we show the improvement in the power spectrum estimation with self-calibration according to three schemes. Even with 1\,s per baseline (corresponding to a full day dedicated to self-calibration), we can reduce significantly the $E\rightarrow B$ leakage. This leakage can be further reduced by spending more time in self-calibration. The three $B$-mode power spectra in black solid lines are the theoretically-expected spectra for three values of the parameter $r$, i.e., the ratio between the amplitudes of the tensor and scalar fluctuations during inflation.

\vspace{-12pt}

\subsection{Spectral Imaging}
\label{sec_spectral_imaging}

  The interferometric nature of QUBIC provides us with another unique feature: the possibility to split the data of each main frequency band into sub-bands, thus considerably increasing the leverage in the control of astrophysical foregrounds. 
  
\begin{figure}[H]
\centering
     \includegraphics[width=15cm]{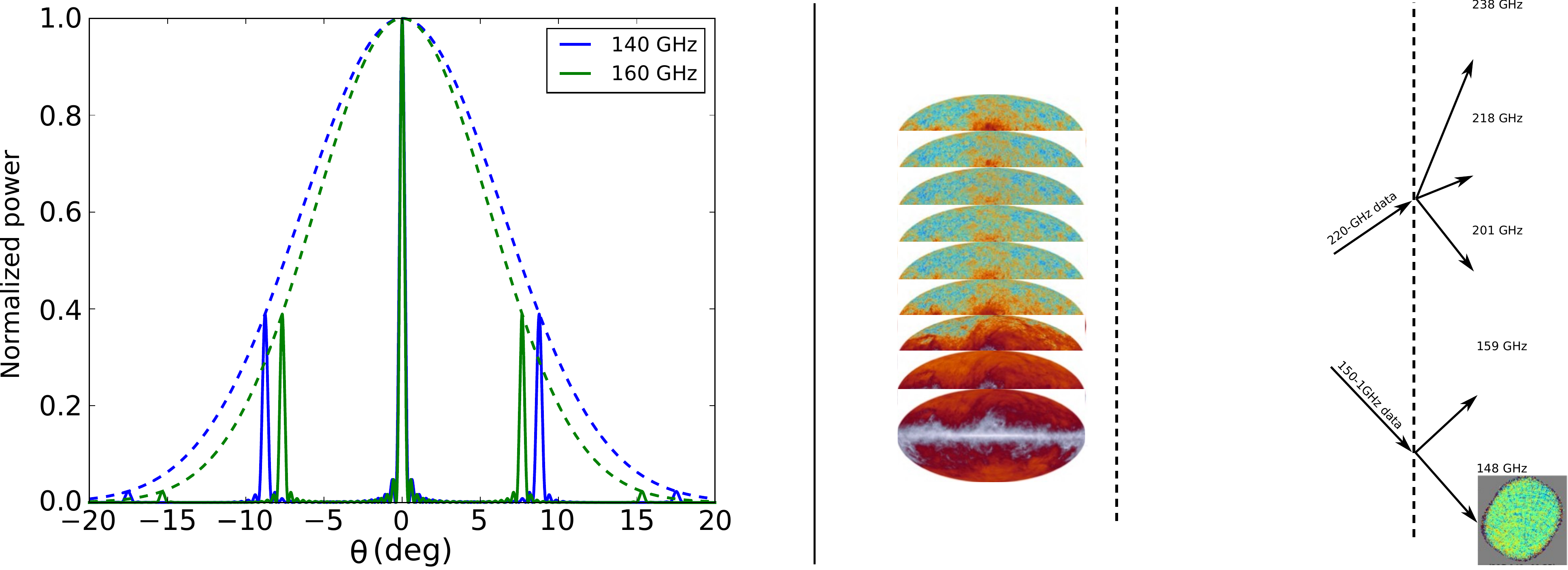}
     \caption{\label{fig_spectral_imaging}({\textbf{Left}}) Cut of the synthesized beam for two monochromatic signals. The main central peaks superimpose each other, but the first lobes are separated so that they can be resolved. The dashed lines represent the horn beams. ({\textbf{Right}}) Schematics representing the ability to resolve spectral sub-bands in each of the main QUBIC bands. The instrument first separates the wide-band sky signal into two main bands (150 and 220\,GHz), then we further separate each band into sub-bands thanks to the spectral sensitivity of the synthesized beams. The Planck maps shown on the left have the only purpose of explaining the concept.}
  \end{figure}

  This feature is called ``spectral imaging'', and its concept is explained schematically in Figure~\ref{fig_spectral_imaging}. The~left panel of Figure~\ref{fig_spectral_imaging} shows the synthetic beams (solid lines) and main feedhorn beams (dashed~lines) at two monochromatic frequencies. The figure clearly shows that the sidelobe peaks are well separated, and this sensitivity of the synthetic beam to the frequency can be exploited to separate the various sub-bands in the input data.

  The right panel of Figure~\ref{fig_spectral_imaging} shows schematically the process of sub-band separation in the data analysis. The instrument measures the wide-band sky signal and splits the two main frequency bands of 150\,GHz and 220\,GHz. Then, the main bands are further separated in the data analysis pipeline by exploiting the frequency sensitivity of the synthetic beam. The interested reader can find further details about the QUBIC data analysis in \cite{Battistelli2011}.
  
  As spectral imaging opens new possibilities of foreground control, it also requires component separation codes that have been not been developed yet. We are currently working on a first detailed assessment of its potential and will soon submit a specific paper on spectral imaging.


\section{The QUBIC Site}
\label{sec_site}

\begin{figure}[H]
\centering
   \includegraphics[width=15cm]{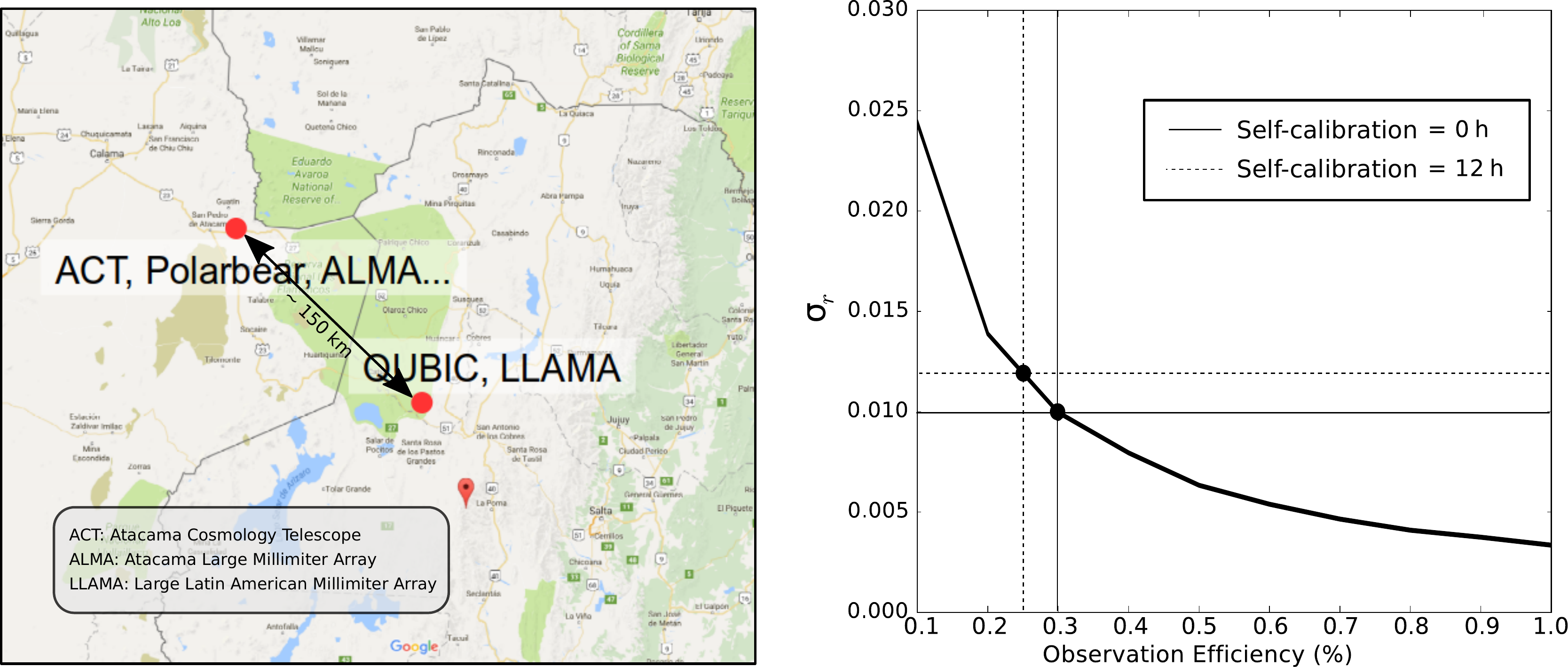}
   \caption{\label{fig_site_performance}({\textbf{Left}}) Location of the QUBIC site compared to the Atacama plateau; ({\textbf{right}}) uncertainty in the tensor-to-scalar ratio, $r$, as a function of the fraction of usable time for two years of operations. The~parameter $r$ was computed considering noise-only simulations.}
\end{figure}

QUBIC will be deployed in Argentina, at the Alto Chorrillos mountain site (24$^\circ$11$'$11.7$''$ S; 66$^\circ$28$'$40.8$''$ W, altitude of 4869 m a.s.l.) near San Antonio de los Cobres, in the Salta province \cite{debernardis2018} (see~the left panel of Figure~\ref{fig_site_performance}). The zenith optical depth measured at 210 GHz, $\tau_{210}$, is <0.1 for 50\% of the year and <0.2 for 85\% of the year. Winds are usually mild ($<6$\,m/s for 50\% of the year), which suggests limited turbulence. 

While the statistics for $\tau_{210}$ in Alto Chorrillos is worse than that of an Antarctic site (either South Pole or Dome C), the site access and logistics are easier. Our trade off is also justified by the following two facts: (i) the atmospheric emission is not polarized to first order, and (ii) a bolometric interferometer intrinsically rejects large-scale atmospheric gradients, which produce most of the atmospheric noise.

The right panel in Figure~\ref{fig_site_performance} (adapted from \cite{Aumont2016}) shows the overall site quality. In the plot, we see the uncertainty in the tensor-to-scalar ratio, $r$, as a function of the fraction of usable time for two years of operations. The circled point shows the estimate for the case in which zero or 12 h are spent each day in calibration mode. The plot shows that for a 30\% usable time (a conservative estimate for our site), we can reach a sensitivity on $r$ of $10^{-2}$ with two years of operations. 

\section{Scientific Performance}
\label{scientific_performance}

In Table~\ref{experiments_summary}, we provide a summary of the expected sensitivity of QUBIC to the {tensor-to-scalar ratio} parameter, $r$, compared with the major ground-based experiments in the same frequency range, either running or expected to be deployed in the near future. 

From the table, we see that all the experiments compete for a detection of $r$ in the range of 10$^{-2}$--10$^{-3}$. The expected sensitivity, however, is not the only performance parameter to be considered in ranking various experiments. Susceptibility to systematic effects and foreground control are as important as the white noise sensitivity. In this respect, QUBIC is unique in the landscape of CMB experiments: it~is based on a completely different design, is less prone to instrumental systematic effects, and allows a~deeper control of foregrounds thanks to its spectral imaging capability.

\begin{table}[H]
\centering
  \caption{Summary of the main $B$-mode ground experiments operating in a frequency range similar to QUBIC. The label ``fg'' or ``no fg'' corresponds to the assumption on the foregrounds; numbers have been extracted from~\cite{Josquin}\label{experiments_summary}.} 
   \begin{tabular}{ccccccc }
     \toprule 
     \multirow{2}{*}{\textbf{Project\vspace{-4pt}
}} &      \multirow{2}{*}{\textbf{Frequencies (GHz)\vspace{-4pt}}} &     \multirow{2}{*}{\boldmath{ $\ell$} \textbf{Range}\vspace{-4pt}} &     \multirow{2}{*}{ \textbf{ Ref.\vspace{-4pt}}} &
     \multicolumn{2}{c}{ \boldmath{$\sigma{(r)}$} \textbf{Goal} } \\ \cmidrule{5-6}
     &      &      &     & \textbf{no fg. }     & \textbf{with fg.} \\ 
     \midrule
     QUBIC     & 150,220     & 30--200 &      &$6.0\times 10^{-3}$ & $1.0\times 10^{-2}$ \\
     Bicep3/Keck  & 95, 150, 220   & 50--250 &\cite{bicep3} &$2.5\times 10^{-3}$ & $1.3\times 10^{-2}$ \\
     CLASS $^\star$
				& 38, 93, 148, 217 & 2--100 &\cite{class} &$1.4\times 10^{-3}$ & $3.0\times 10^{-3}$ \\
     SPT-3G $^\dagger$
				& 95, 148, 223   & 50--3000 &\cite{spt3g} &$1.7\times 10^{-3}$ & $5.0\times 10^{-3}$ \\
     AdvACT $^\ddagger$
				& 90, 150, 230   & 60--3000 &\cite{advact} &$1.3\times 10^{-3}$ & $4.0\times 10^{-3}$ \\
     Simons Array  & 90, 150, 220   & 30--3000 &\cite{simons} &$1.6\times 10^{-3}$ & $5.0\times 10^{-3}$ \\
     \bottomrule
   \end{tabular}
   \begin{tabular}{@{}c@{}} 
\multicolumn{1}{p{\textwidth -.88in}}{\footnotesize $^\star$ CLASS: Cosmology Large Angular Scale Surveyor; $^\dagger$ SPT-3G: South Pole Telescope---3rd generation; $^\ddagger$~AdvACT: Advanced Atacama Cosmology Telescope.}
\end{tabular}
\end{table}


\section{Current Status}
\label{sec_current_status}

\textls[-20]{QUBIC is currently in the phase of laboratory calibration of the so-called ``Technological Demonstrator''} (TD). The TD has a reduced focal plane and horn array with respect to the full instrument. In particular, the TD has only one-quarter of the 150\,GHz TES focal plane, an array of $64+64$ horns, 64 switches, and a smaller optical combiner. The TD will not produce science, but it will demonstrate the feasibility of the bolometric interferometry both in the laboratory and in the field.

\begin{figure}[H]
\centering
   \includegraphics[width = 14cm]{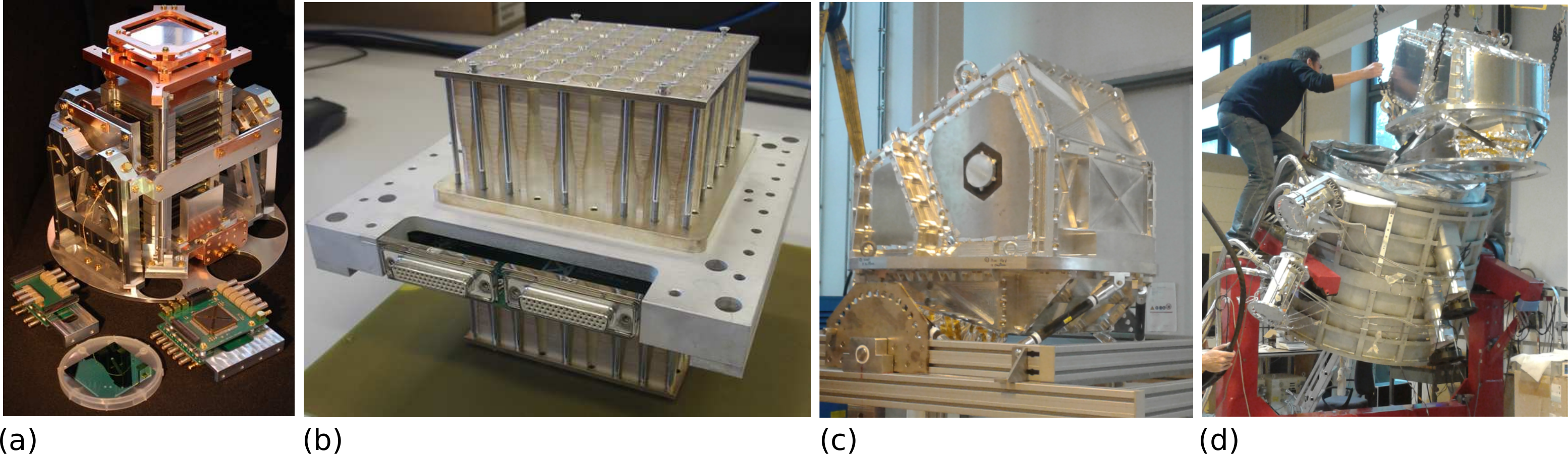}
   \caption{\label{fig_instrument_calibration}Status of the current QUBIC development. (\textbf{a}) The cryogenic section of the QUBIC detection chain; (\textbf{b}) the TD array of $64+64$ back-to-back horns interfaced with their switches; (\textbf{c}) the integrated 1\,K box; (\textbf{d}) integration of the 1\,K 
	box in the cryostat shell.}
\end{figure}

In Figure~\ref{fig_instrument_calibration}, we show various QUBIC components. Panel (a) shows one of the two cryogenic detection chains. On top of the chain, one can see the TES focal plane. Panel (b) shows the array of the $64+64$ back-to-back dual-band corrugated horns interfaced with the switch array. Panels (c) and (d) show the 1\,K box before and during the integration into the QUBIC cryostat.

As of November 2018, QUBIC TD is cold and being calibrated at the Laboratoire AstroParticule \& Cosmologie (APC) in Paris. This testing phase will end at the beginning of Spring 2019, when QUBIC will be shipped to Argentina and installed at the site for a first-light test foreseen within 2019. In the meantime, in Europe, we will proceed to the fabrication of the final instrument missing parts: the full TES focal planes, the array of $400+400$ horns interfaced with the 400 switches, and the final optical combiner. The deployment of QUBIC final instrument is foreseen to be completed by 2020.


\section{Conclusions}
\label{sec_conclusions}

QUBIC is a new way to measure the polarization of the CMB. It combines the sensitivity of TES bolometric arrays with the control of systematic effects that are typical of interferometers. This is a key asset in CMB polarization experiments, where high sensitivity must be backed by comparable levels of systematic effects and foreground control. QUBIC responds to this challenge with the key features of self-calibration and spectral imaging, which are possible thanks to the interferometric nature of the instrument. A technological demonstrator is currently being tested in the laboratory and will soon be deployed in Argentina for a first-light test. We forecast the installation of the final instrument and the start of scientific operations during 2020, opening the way for a new generation of instruments in the field of Cosmic Microwave Background polarimetry.

\vspace{6pt} 


\funding{QUBIC is funded by the following agencies. {France}: ANR (Agence Nationale de la Recherche) 2012 and 2014, DIM-ACAV (Domaine d'Interet Majeur---Astronomie et Conditions d'Apparition de la Vie), CNRS/IN2P3 (Centre national de la recherche scientifique/Institut national de physique nucléaire et de physique des particules), CNRS/INSU (Centre national de la recherche scientifique/Institut national de sciences de l'univers). Maria Salatino acknowledges the financial support of the UnivEarthS Labex program at Sorbonne Paris Cité (ANR-10-LABX-0023 and ANR-11-IDEX-0005-02. {Italy}: CNR/PNRA (Consiglio Nazionale delle Ricerche/Programma Nazionale Ricerche in Antartide) until 2016, INFN (Istituto Nazionale di Fisica Nucleare) since 2017. {Argentina}: Secretaría de Gobierno de Ciencia, Tecnología e Innovación Productiva, Comisión Nacional de Energía Atómica, Consejo Nacional de Investigaciones Científicas y Técnicas. {U.K.}: the University of Manchester team acknowledges the support of STFC (Science and Technology Facilities Council) grant ST/L000768/1. {Ireland}: James Murphy and David Burke acknowledge postgraduate scholarships from the Irish Research Council. {Duc Hoang Thuong} acknowledges the Vietnamese government for funding his scholarship at APC. {Andrew May} acknowledges the support of an~STFC PhD Studentship.}


\conflictsofinterest{The authors declare no conflicts of interest.}




%

\begin{thebibliography}{-------}
\providecommand{\natexlab}[1]{#1}

\bibitem[Battistelli \em{et~al.}(2011)Battistelli, Ba{\'{u}}, Bennett,
  Berg{\'{e}}, Bernard, de~Bernardis, Bordier, and et~al]{Battistelli2011}
Battistelli, E.; Ba{\'{u}}, A.; Bennett, D.; Berg{\'{e}}, L.; Bernard, J.P.;
  de~Bernardis, P.; Bordier, G.; et~al.
\newblock {QUBIC: The QU bolometric interferometer for cosmology}.
\newblock {\em Astroparticle Physics} {\bf 2011}, {\em 34},~705--716.
\newblock
  doi:{\changeurlcolor{black}\href{https://doi.org/10.1016/j.astropartphys.2011.01.012}{\detokenize{10.1016/j.astropartphys.2011.01.012}}}.

\bibitem[Tartari \em{et~al.}(2015)Tartari, Aumont, Banfi, Battaglia,
  Battistelli, Ba{\`{u}}, B{\'{e}}lier, and et~al.]{qubic15}
Tartari, A.; Aumont, J.; Banfi, S.; Battaglia, P.; Battistelli, E.S.;
  Ba{\`{u}}, A.; B{\'{e}}lier, B.; et~al..
\newblock {QUBIC: A Fizeau Interferometer Targeting Primordial B-Modes}.
\newblock {\em Journal of Low Temperature Physics} {\bf 2015}, {\em 181}.
\newblock
  doi:{\changeurlcolor{black}\href{https://doi.org/10.1007/s10909-015-1398-3}{\detokenize{10.1007/s10909-015-1398-3}}}.

\bibitem[Aumont \em{et~al.}(2016)Aumont, Banfi, Battaglia, Battistelli,
  Ba{\`{u}}, and et~al]{Aumont2016}
Aumont, J.; Banfi, S.; Battaglia, P.; Battistelli, E.S.; Ba{\`{u}}, A.; et~al.
\newblock {QUBIC Technical Design Report}.
\newblock {\em astro-ph pre-prints} {\bf 2016},
  \href{http://xxx.lanl.gov/abs/1609.04372}{{\normalfont [1609.04372]}}.

\bibitem[Liu \em{et~al.}(2010)Liu, Tegmark, Morrison, Lutomirski, and
  Zaldarriaga]{liu2010}
Liu, A.; Tegmark, M.; Morrison, S.; Lutomirski, A.; Zaldarriaga, M.
\newblock Precision calibration of radio interferometers using redundant
  baselines.
\newblock {\em Monthly Notices of the Royal Astronomical Society} {\bf 2010},
  {\em 408},~1029--1050.
\newblock
  doi:{\changeurlcolor{black}\href{https://doi.org/10.1111/j.1365-2966.2010.17174.x}{\detokenize{10.1111/j.1365-2966.2010.17174.x}}}.

\bibitem[Bigot-Sazy \em{et~al.}(2013{\natexlab{a}})Bigot-Sazy, Charlassier,
  Hamilton, Kaplan, and Zahariade]{Bigot-Sazy2013}
Bigot-Sazy, M.A.; Charlassier, R.; Hamilton, J.; Kaplan, J.; Zahariade, G.
\newblock {Self-calibration: an efficient method to control systematic effects
  in bolometric interferometry}.
\newblock {\em Astronomy {\&} Astrophysics} {\bf 2013}, {\em 59},~1--11.
\newblock
  doi:{\changeurlcolor{black}\href{https://doi.org/10.1051/0004-6361/201220429}{\detokenize{10.1051/0004-6361/201220429}}}.

\bibitem[Bigot-Sazy \em{et~al.}(2013{\natexlab{b}})Bigot-Sazy, Charlassier,
  Hamilton, Kaplan, and Zahariade]{bs13}
Bigot-Sazy, M.A.; Charlassier, R.; Hamilton, J.; Kaplan, J.; Zahariade, G.
\newblock {Self-calibration: an efficient method to control systematic effects
  in bolometric interferometry}.
\newblock {\em Astronomy {\&} Astrophysics} {\bf 2013}, {\em 59},~1--11.
\newblock
  doi:{\changeurlcolor{black}\href{https://doi.org/10.1051/0004-6361/201220429}{\detokenize{10.1051/0004-6361/201220429}}}.

\bibitem[{de Bernardis} \em{et~al.}(2018){de Bernardis}, {Ade}, {Amico},
  {Auguste}, {Aumont}, and et~al.]{debernardis2018}
{de Bernardis}, P.; {Ade}, P.; {Amico}, G.; {Auguste}, D.; {Aumont}, J.;
  et~al..
\newblock {QUBIC: Measuring CMB polarization from Argentina}.
\newblock {\em Boletin de la Asociacion Argentina de Astronomia La Plata
  Argentina} {\bf 2018}, {\em 60},~107--114.

\bibitem[Hui \em{et~al.}(2016)Hui, Ade, Ahmed, Alexander, Amiri, Barkats,
  Benton, and et~al.]{bicep3}
Hui, H.; Ade, P.A.R.; Ahmed, Z.; Alexander, K.D.; Amiri, M.; Barkats, D.;
  Benton, S.J.; et~al..
\newblock BICEP3 focal plane design and detector performance.
\newblock  SPIE proceedings for Millimeter, Submillimeter, and Far-Infrared
  Detectors and Instrumentation for Astronomy VIII,  2016, Vol. 9914, pp. 9914
  -- 9914 -- 11.
\newblock
  doi:{\changeurlcolor{black}\href{https://doi.org/10.1117/12.2232986}{\detokenize{10.1117/12.2232986}}}.

\bibitem[{Harrington} \em{et~al.}(2016){Harrington}, {Marriage}, {Ali},
  {Appel}, {Bennett}, {Boone}, {Brewer}, and et~al.]{class}
{Harrington}, K.; {Marriage}, T.; {Ali}, A.; {Appel}, J.W.; {Bennett}, C.L.;
  {Boone}, F.; {Brewer}, M.; et~al..
\newblock {The Cosmology Large Angular Scale Surveyor}.
\newblock  Millimeter, Submillimeter, and Far-Infrared Detectors and
  Instrumentation for Astronomy VIII,  2016, Vol. 9914, {\em Society of
  Photo-Optical Instrumentation Engineers (SPIE) Conference Series}, p. 99141K,
   \href{http://xxx.lanl.gov/abs/1608.08234}{{\normalfont
  [arXiv:astro-ph.IM/1608.08234]}}.
\newblock
  doi:{\changeurlcolor{black}\href{https://doi.org/10.1117/12.2233125}{\detokenize{10.1117/12.2233125}}}.

\bibitem[{Benson} \em{et~al.}(2014){Benson}, {Ade}, {Ahmed}, {Allen}, {Arnold},
  {Austermann}, {Bender}, and et~al.]{spt3g}
{Benson}, B.A.; {Ade}, P.A.R.; {Ahmed}, Z.; {Allen}, S.W.; {Arnold}, K.;
  {Austermann}, J.E.; {Bender}, A.N.; et~al..
\newblock {SPT-3G: a next-generation cosmic microwave background polarization
  experiment on the South Pole telescope}.
\newblock  Millimeter, Submillimeter, and Far-Infrared Detectors and
  Instrumentation for Astronomy VII,  2014, Vol. 9153, {\em Society of
  Photo-Optical Instrumentation Engineers (SPIE) Conference Series}, p. 91531P,
   \href{http://xxx.lanl.gov/abs/1407.2973}{{\normalfont
  [arXiv:astro-ph.IM/1407.2973]}}.
\newblock
  doi:{\changeurlcolor{black}\href{https://doi.org/10.1117/12.2057305}{\detokenize{10.1117/12.2057305}}}.

\bibitem[{Li} \em{et~al.}(2018){Li}, {Austermann}, {Beall}, {Bruno}, {Choi},
  {Cothard}, {Crowley}, and et~al.]{advact}
{Li}, Y.; {Austermann}, J.E.; {Beall}, J.A.; {Bruno}, S.M.; {Choi}, S.K.;
  {Cothard}, N.F.; {Crowley}, K.T.; et~al..
\newblock {Performance of the advanced ACTPol low frequency array}.
\newblock  Society of Photo-Optical Instrumentation Engineers (SPIE) Conference
  Series,  2018, Vol. 10708, p. 107080A.
\newblock
  doi:{\changeurlcolor{black}\href{https://doi.org/10.1117/12.2313942}{\detokenize{10.1117/12.2313942}}}.

\bibitem[{Keating}(2018)]{simons}
{Keating}, B.
\newblock {The POLARBEAR and Simons Array CMB Polarization Experiments}.
\newblock  42nd COSPAR Scientific Assembly,  2018, Vol.~42, pp. E1.2--25--18.

\bibitem[{Chapman} \em{et~al.}(2014){Chapman}, {Aboobaker}, {Ade}, {Aubin},
  {Baccigalupi}, {Bandura}, {Bao}, and et~al.]{ebex10k}
{Chapman}, D.; {Aboobaker}, A.M.; {Ade}, P.; {Aubin}, F.; {Baccigalupi}, C.;
  {Bandura}, K.; {Bao}, C.; et~al..
\newblock {EBEX: A Balloon-Borne CMB Polarization Experiment}.
\newblock  American Astronomical Society Meeting Abstracts \#223,  2014, Vol.
  223, {\em American Astronomical Society Meeting Abstracts}, p. 407.03.

\bibitem[{Gualtieri} \em{et~al.}(2018){Gualtieri}, {Filippini}, {Ade}, {Amiri},
  {Benton}, {Bergman}, {Bihary}, {Bock}, {Bond}, {Bryan}, and {Chiang}]{spider}
{Gualtieri}, R.; {Filippini}, J.P.; {Ade}, P.A.R.; {Amiri}, M.; {Benton}, S.J.;
  {Bergman}, A.S.; {Bihary}, R.; {Bock}, J.J.; {Bond}, J.R.; {Bryan}, S.A.;
  {Chiang}, e.a.
\newblock {SPIDER: CMB Polarimetry from the Edge of Space}.
\newblock {\em Journal of Low Temperature Physics} {\bf 2018}, {\em
  193},~1112--1121,  \href{http://xxx.lanl.gov/abs/1711.10596}{{\normalfont
  [arXiv:astro-ph.CO/1711.10596]}}.
\newblock
  doi:{\changeurlcolor{black}\href{https://doi.org/10.1007/s10909-018-2078-x}{\detokenize{10.1007/s10909-018-2078-x}}}.

\bibitem[{Lazear} \em{et~al.}(2014){Lazear}, {Ade}, {Benford}, {Bennett},
  {Chuss}, {Dotson}, {Eimer}, and et~al.]{piper}
{Lazear}, J.; {Ade}, P.A.R.; {Benford}, D.; {Bennett}, C.L.; {Chuss}, D.T.;
  {Dotson}, J.L.; {Eimer}, J.R.; et~al..
\newblock {The Primordial Inflation Polarization Explorer (PIPER)}.
\newblock  Millimeter, Submillimeter, and Far-Infrared Detectors and
  Instrumentation for Astronomy VII,  2014, Vol. 9153, {\em Society of
  Photo-Optical Instrumentation Engineers (SPIE) Conference Series}, p. 91531L,
   \href{http://xxx.lanl.gov/abs/1407.2584}{{\normalfont
  [arXiv:astro-ph.IM/1407.2584]}}.
\newblock
  doi:{\changeurlcolor{black}\href{https://doi.org/10.1117/12.2056806}{\detokenize{10.1117/12.2056806}}}.

\bibitem[Errard \em{et~al.}(2016)Errard, Feeney, Peiris, and Jaffe]{Josquin}
Errard, J.; Feeney, S.M.; Peiris, H.V.; Jaffe, A.H.
\newblock Robust forecasts on fundamental physics from the foreground-obscured,
  gravitationally-lensed CMB polarization.
\newblock {\em Journal of Cosmology and Astroparticle Physics} {\bf 2016}, {\em
  2016},~052.

\end{thebibliography}
\reftitle{References}

\end{document}